\documentclass{article}

     \PassOptionsToPackage{numbers, compress}{natbib}
     \usepackage[preprint]{neurips_2020}

\usepackage[utf8]{inputenc} % allow utf-8 input
\usepackage[T1]{fontenc}    % use 8-bit T1 fonts
\usepackage[hidelinks]{hyperref}       % hyperlinks
\usepackage{url}            % simple URL typesetting
\usepackage{booktabs}       % professional-quality tables
\usepackage{amsfonts}       % blackboard math symbols
\usepackage{nicefrac}       % compact symbols for 1/2, etc.
\usepackage{microtype}      % microtypography
\usepackage{natbib}
\usepackage{graphicx}
\usepackage{multicol}
\usepackage{multirow}
\usepackage[dvipsnames]{xcolor}

\title{Evaluating the Ebb and Flow: An In-depth Analysis of Question-Answering Trends across Diverse Platforms}

\author{
Rima Hazra, Agnik Saha, Somnath Banerjee, Animesh Mukherjee\\
Indian Institute of Technology Kharagpur\\ 
\texttt{ to\_rima@iitkgp.ac.in}\\
\texttt{ som.iitkgpcse@kgpian.iitkgp.ac.in, animeshm@cse.iitkgp.ac.in} 
}

\begin{document}

\maketitle

\begin{abstract}
  % Community question answering (CQA) platforms are becoming increasingly popular among the users for obtaining quick response to their queries. Quick response of the queries are dependent on various query as well as user related factors. In this paper, we investigate these factors for six CQA platforms chosen based on their popularity (i.e., \textit{answering speed}). We observe that the metadata, the structure of the questions and the interaction among the users are correlated with the time required to receive the first answer for a question. Further, we use standard machine learning models on these metadata and interaction among users to predict the questions getting first answer quickly.
  Community Question Answering (CQA) platforms steadily gain popularity as they provide users with fast responses to their queries. The swiftness of these responses is contingent on a mixture of query-specific and user-related elements. This paper scrutinizes these contributing factors within the context of six highly popular CQA platforms, identified through their standout ``\textit{answering speed}". Our investigation reveals a correlation between the time taken to yield the first response to a question and several variables: the metadata, the formulation of the questions, and the level of interaction among users. Additionally, by employing conventional machine learning models to analyze these metadata and patterns of user interaction, we endeavor to predict which queries will receive their initial responses promptly.

% \keywords{community question answering \and response time \and asker-answerer graph \and classification model}
\end{abstract}
\section{Introduction}
Community question-answering platforms are gradually evolving over the past decade. Such portals allow users to post their queries and answer the questions. These CQA platforms help experts share their knowledge and new users solve their queries. Initially, these community question-answering platforms focused on providing the most relevant answers to the users' queries~\cite{Zhang_2:2021}\cite{Bachschi:2020}. As the CQA platforms keep growing, they not only focus on providing the relevant answers but also on quickly delivering such answers~\cite{Yazdaninia:2021}\cite{Wang:2013}. The growth and maturity of such platforms is contingent on a number of variables, including the ``quality of questions and answers", the ``response time of queries", ``user involvement", and ``user activity"~\cite{Moutidis:2021}. The users with expertise in specific domains significantly help the platform's growth. Based on the users' activeness on the platform and the importance of their posts, they have been rewarded reputation scores, badges, and additional privileges. In most such CQA platforms driven by a reputation system, the platform's popularity depends on the response time~\cite{Bhat:2014} of the question (i.e., time to receive the first answer). So, the users who want to post a query may also want to know the time by which they can expect a response to their question. StackExchange\footnote{https://stackexchange.com/} is one of the CQA platforms where the users can ask questions and receive answers from other users. % On the StackExchange website, multiple CQA websites focus on the queries related to Mathematics, English, Software Engineering, Money, Chemistry, and Game Development. 
An analysis of such a platform's dynamics helps the community maintainers adapt and design the framework to address various CQA-related problems~\cite{Anderson:2012}~\cite{Mondal:2021}~\cite{Zhang:2021}~\cite{Wang:2018}~\cite{hazra2024duplicate}~\cite{10.1007/978-3-031-26422-1_15}~\cite{10.1007/978-3-030-72113-8_15}. In this paper, we examine the dynamics of different CQA platforms relating the usual response time with other factors of the portals. We conduct an empirical study on six CQA platforms -- Mathematics, Software Engineering, English, Game Development, Chemistry, and Money. To analyze the dynamics of the diverse community question-answering platforms, we look into various factors mentioned here -- 
(1) Characterize the questions in terms of textual information, linguistic characteristics, question tags, votes received by the questions etc. We obtain correlations of these features with the time required to respond to a question. %We conduct an analysis on response time of the question, getting accepted answers within tangible time. Further we try to related the question centric factors with the response time of the questions.   %Analyzing the platforms in terms of the questions such as questions textual information, its tags. Further, we try to relate these factors to the response time of the questions. 
(2) Characterizing the users of the platform in terms of their reputation \& activity in the platforms. Further, we build Asker-Answerer graph (AAG) where the nodes are the users and a link is created if the question from a user is answered by another user. Subsequently, we conduct in depth analysis of the characteristics of the common users present across all the platforms.
%    \item With the help of the AAG, we observe the transition of the common users from the outer core to the innermost cores in the graph over time.
 (3)For all the platforms, we use metadata of the questions, question structure and user interaction as features to build a standard machine learning model for classifying the newly posted questions which are having fast answer.
%\vspace*{-0.2cm}
% In previous works, everyone has found correlations between question properties with response time. Our novelty was showing how network features of our asker and answerer graph effect response time and some interesting user findings (cross platform users, etc.) and ml classifiers distinguish high- and low-response questions
% In most of the earlier works~\cite{10.1145/3180155.3182521}, the correlation between various question-related factors and the response time were examined. In this study, our novelty lies in understanding how various underlying network (asker-answerer graph) related features along with the question related features are related to the response time.
The majority of preceding studies~\cite{10.1145/3180155.3182521} scrutinize the correlation between diverse question-related factors and response time. The novelty of our current research resides in our endeavor to comprehend how various inherent network-related features (reflected in the asker-answerer graph) coincide with question-related features in relation to response time.
In this study, we make the following observations --
%\vspace*{-0.2cm}
%\begin{enumerate}
    \noindent (1) \textit{Short questions get faster responses.} Chemistry and Software Engineering questions are harder. \textit{More tags mean slower replies.} Math, Software Engineering, and English answers come in an hour, others in a day. In Software Engineering and English, many questions get $\ge 3$ answers.%A question body with less number of words always has a better response time than others. Questions from Chemistry and Software Engineering are harder to read. The number of tags used in a question are correlated with response time. Domains like Mathematics, Software Engineering and English on average receive answers within an hour while for others it takes 24 hours. %Some domains like Mathematics, Software Engineering and English receive answers pretty quickly on average (within an hour from the post of the question). For other domains this timegap could be as high as 24 hours. %But for other platforms, this timegap become 24 hours. 
    %A significant fraction of questions receive $\ge 3$ answers in Software Engineering and English domains. 
    \noindent (2) Most users (~60\%-70\%) on all platforms stay inactive. Domains like Money, Game Development, and Chemistry have more active users than Mathematics, Software Engineering, and English. %This possibly indicates that in the former domains questions change very fast propelling new question posts which naturally would require newer answers. For the latter domains however many of the answers could be already found existing on the CQA platform. %Such pattern infer that users in Money, Game Development and Chemistry may have very specific questions and may not get already existing appropriate answers from the platform. In case of Mathematics, Software Engineering and English, users are able resolve their queries by looking at the existing one.
    \noindent (3) XGBoost and MLP perform best at classifying new questions on most platforms. The second best is usually the random forest model.
%\end{enumerate}
%\vspace*{-0.55cm}
\section{Dataset}
%\vspace*{-0.25cm}
We test six CQA datasets from StackExchange up to March 2022, picking six platforms (top three and bottom three in answering speed) from an initial group that includes Physics, Chemistry, Mathematics, English, Software Engineering, Game Development, AskUbuntu, Mathematica, Travel, Money. Answering speed, different from response time, is the \% of questions answered within time $t$ ($t$ = 10 mins, 20 mins, etc.). Math, English, and Software Engineering are top in speed; Money and Finance, GameDev, and Chemistry are bottom. Data includes question title, body, tags, reporting time, answers with timestamps, votes, and user reputations. Table~\ref{tab:basicStat} presents dataset stats.
\begin{table}[h]
\centering
\tiny
\begin{tabular}{|c|c|c|c|c|} \hline
 Datasets & \#Questions & \#Unique Tags & \#Unique Users & Age (in yr.)\\ 
 \hline
 \em{SE} & 60,801 & 1667 & 3,46,642 & 12 (2010)  \\  \hline
 \em{GD} & 53,427 & 1094 & 1,22,553 & 12 (2010) \\  \hline
 \em{MA} & 14,79,363 & 1898 & 8,88,141 & 12 (2010) \\  \hline
 \em{EN} & 1,24,373 & 981 & 3,50,766 & 12 (2010) \\  \hline
\em{CH} & 40,742 & 368 & 88,677 & 10 (2012)\\  \hline
 \em{MO} & 35,494 & 1005 & 84,650 & 13 (2009)\\  \hline%[1ex]
\end{tabular}
\caption{Basic statistics of six community question answering platforms.}
\label{tab:basicStat}
%\vspace*{-0.75cm}
\end{table}
% \begin{table}[h]
% \resizebox{\textwidth}{!}{%  
% \begin{tabular}{|c|c|c|c|c|c|}%{||T{0.278\textwidth}|T{0.2\textwidth}|T{0.2\textwidth}|T{0.2\textwidth}|T{0.2\textwidth}|T{0.2\textwidth}||}
%  \hline
%  Datasets & Average number  & Total number & \% of answered & \% of questions  & Average number \\ 
%  & of tags & of answers &  questions & having accepted answer & of votes \\ 
%  \hline
%  \em{SE} & 2.802 & 1,72,275 & 95.44 & 57.78 &  6.6139 \\ \hline
%  \em{GD} & 2.795 & 77,996 & 86.31 & 52.80 & 2.567 \\ \hline
%  \em{MA}  & 2.369 & 19,76,51 & 82.32 & 52.71 & 2.11 \\ \hline
%  \em{EN} & 2.08 & 2,79,346 & 92.12 & 48.1 & 3.36 \\ \hline
% \em{CH} & 2.396 & 47,420 & 79.6 & 40.63 & 3.42\\ \hline
%  \em{MO} & 3.11 & 66,917 & 91.19 & 45.32 & 4.71 \\ 
%  \hline
% \end{tabular}
% }
% \caption{Basic statistics about the question structure}
% \label{table:3}
% \end{table}
%\vspace*{-0.85cm}
\begin{table}[h]
\centering
\tiny
\resizebox{\textwidth}{!}{%  
\begin{tabular}{|c|c|c|c|c|c|}%{||T{0.278\textwidth}|T{0.2\textwidth}|T{0.2\textwidth}|T{0.2\textwidth}|T{0.2\textwidth}|T{0.2\textwidth}||}
 \hline
 Datasets & Average number  & Total number & \% of answered & \% of questions  & Average number \\ 
 & of tags & of answers &  questions & having accepted answer & of votes \\ 
 \hline
 \em{SE} & 2.802 & 1,72,275 & 95.44 & 57.78 &  6.6139 \\  \hline
 \em{GD} & 2.795 & 77,996 & 86.31 & 52.80 & 2.567 \\ \hline
 \em{MA}  & 2.369 & 19,76,51 & 82.32 & 52.71 & 2.11 \\ \hline
 \em{EN} & 2.08 & 2,79,346 & 92.12 & 48.1 & 3.36 \\ \hline
\em{CH} & 2.396 & 47,420 & 79.6 & 40.63 & 3.42\\ \hline
 \em{MO} & 3.11 & 66,917 & 91.19 & 45.32 & 4.71 \\ 
 \hline
\end{tabular}
}
\caption{Basic statistics about the question structure}
\label{table:3}
%\vspace*{-0.95cm}
\end{table}
%\vspace*{-0.65cm}
\section{Empirical analysis}
In order to compare the platforms we consider two different factors -- (i) the characteristic features of the questions posted in a platform, and (ii) the interaction behaviour of users on a platform. In each case we correlate these factors with the response time, which is the time required to obtain the first answer after a question is posted.%In order to understand the dynamics of the platform, we mainly focus on the structure of the questions, linguistic characteristics, probability of receiving answers, what kind of questions take longer time to get answered, how the tags are associated with questions. Further we attempt to correlate these factors with the response time of the questions in the platform.  \rh{In order to understand the dynamics of the users, we investigate the characteristics of users interaction in the platform}\am{not clear?}\rh{[added]}. Further, we conduct in depth analysis on the common users across all the platforms. 
%\vspace*{-0.35cm}
\subsection{Characterizing the questions}
\begin{figure*}[ht]
\centering
\small
\includegraphics[height=2.8cm, width=5.3cm]{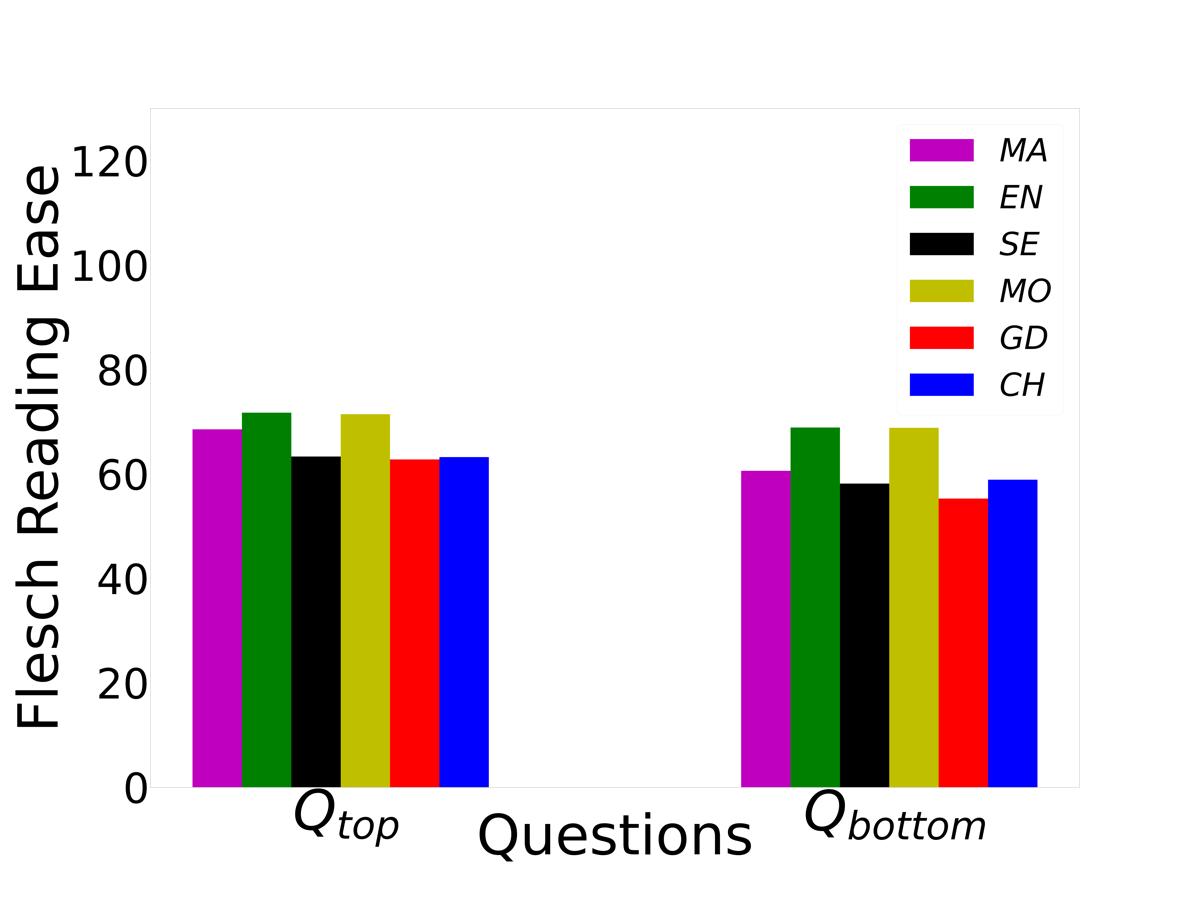} \hfill
\includegraphics[height=2.8cm, width=5.3cm]{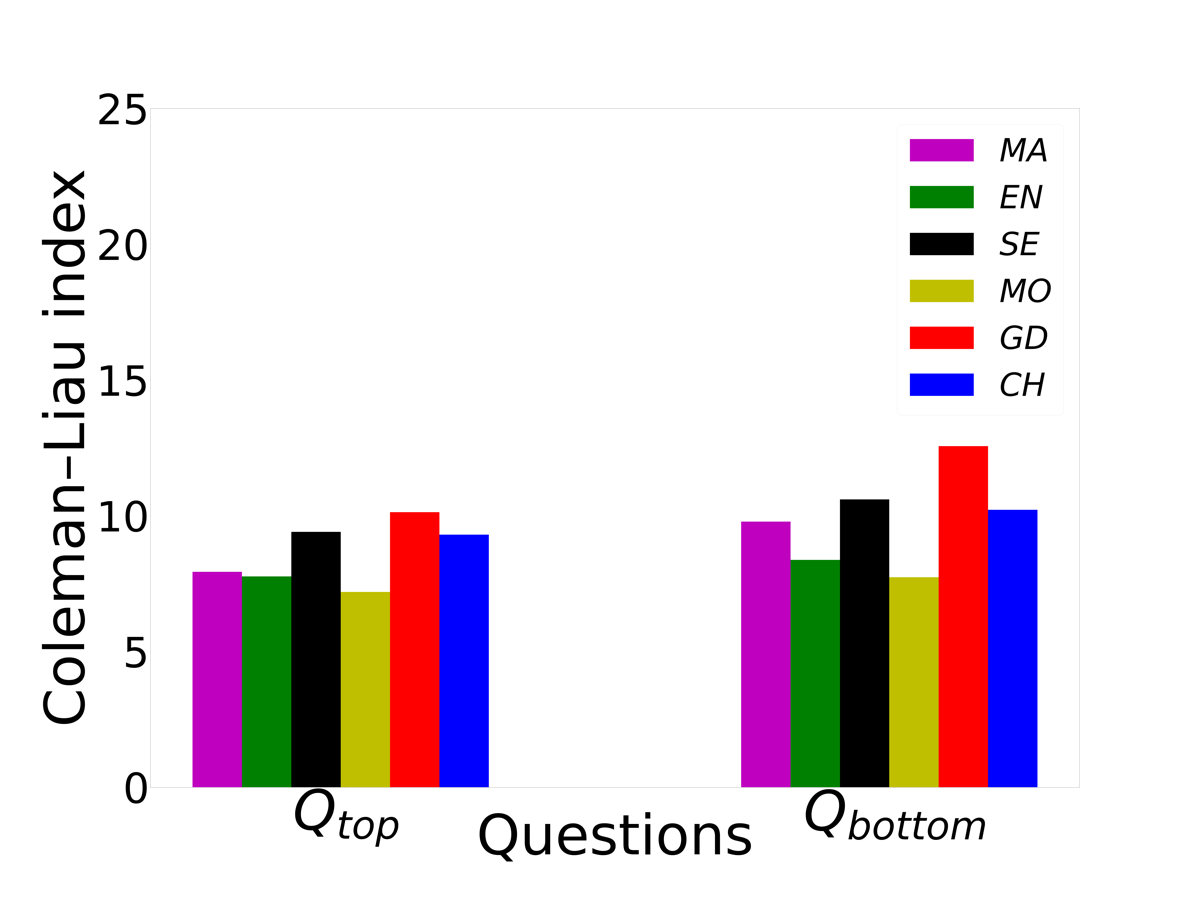} \hfill
\includegraphics[height=2.8cm, width=5.3cm]{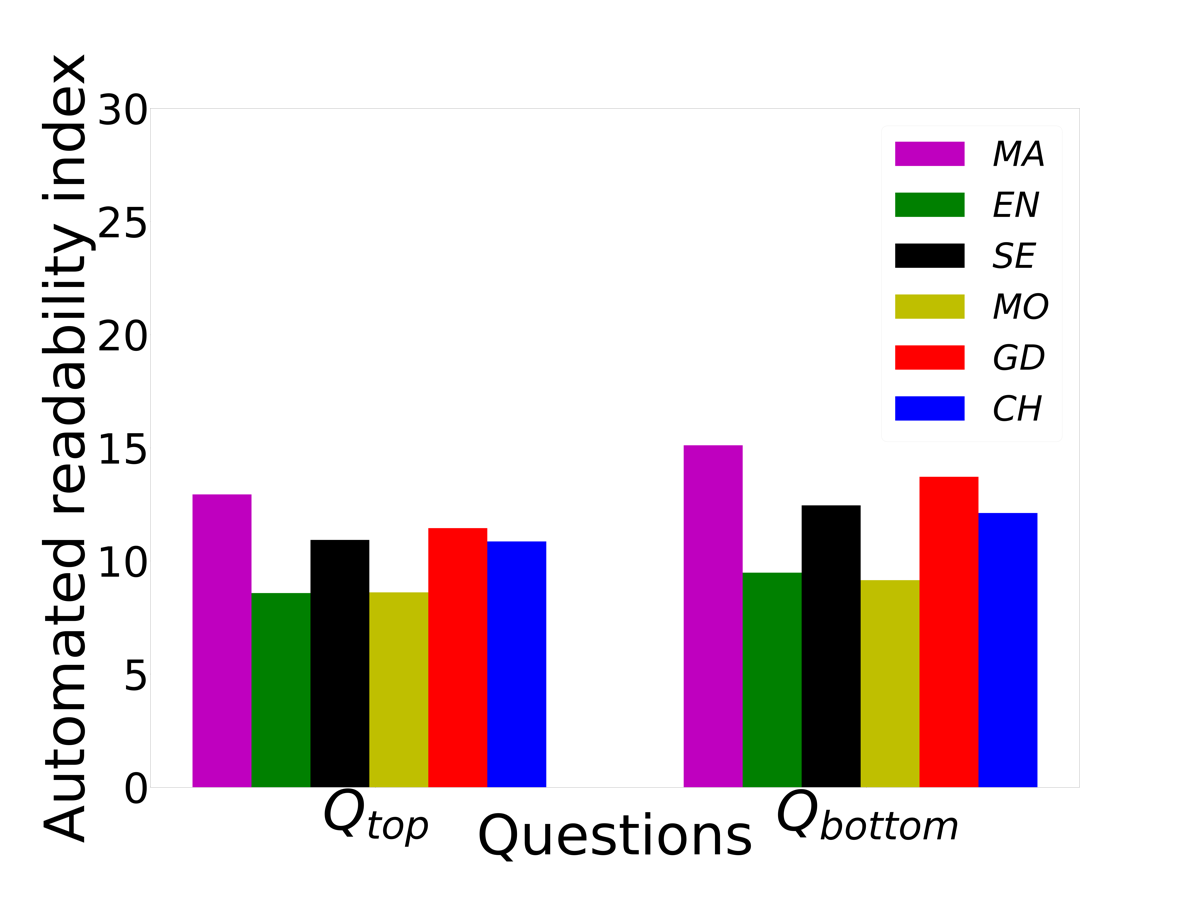} \hfill
\includegraphics[height=2.8cm, width=5.3cm]{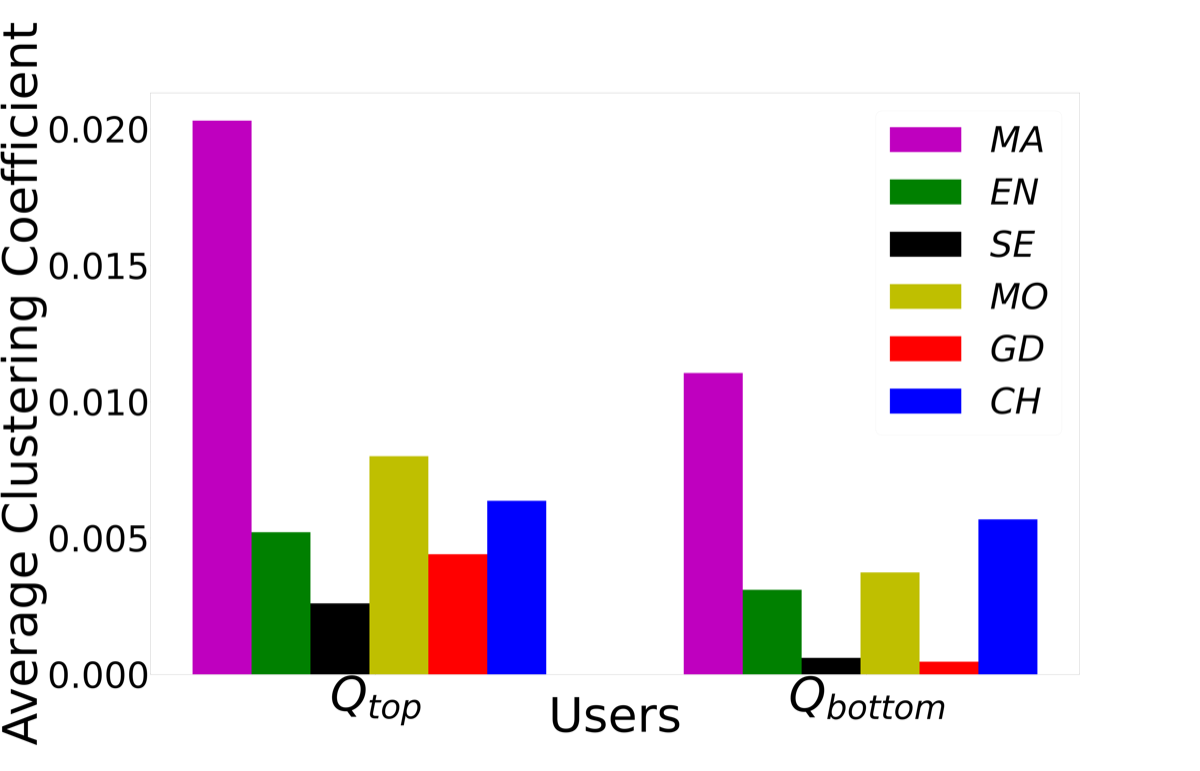} \hfill
\includegraphics[height=2.8cm, width=5.3cm]{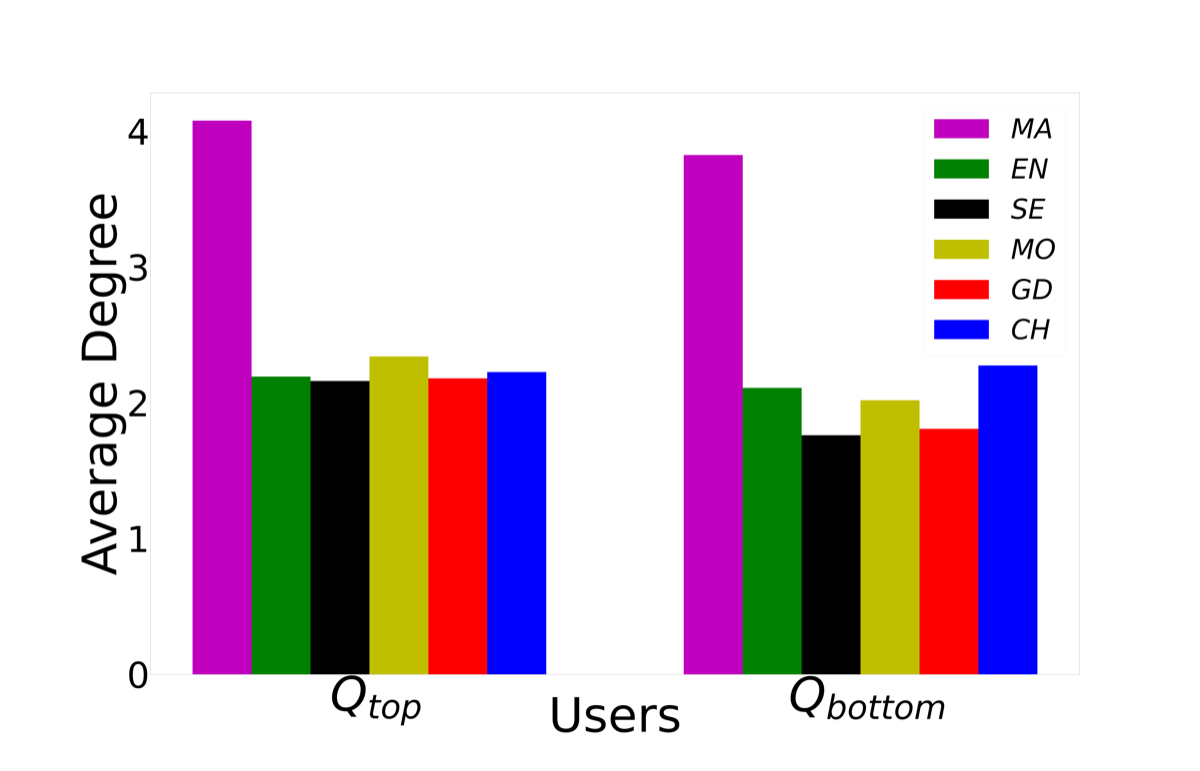} \hfill
\caption{(left): a. Flesch Reading Ease, ($2^{nd}$ from left): b. Coleman-Liau index, (middle): c. Automated readability index scores for $Q_{top}$ and $Q_{bottom}$ questions. ($2^{nd}$ from right): d. Average Clustering Coefficient, (right): e. Average Degree of AAG of $Q_{top}$ and $Q_{bottom}$.}  
\label{fig:readabilityGraphprop}
%\vspace*{-0.85cm}
\end{figure*}

In order to characterize the questions of different platforms, we explore various structural properties of the question. These include length of the title, length of the body, linguistic characteristics of question title and question body. In Table~\ref{table:3}, we have shown the average number of tags present in the question, total number of answers, percentage of questions having at least one answer, percentage of questions having accepted answer, average number of votes received by questions. From Table~\ref{table:3}, it is observed that 90\%-95\% of questions have received at least one answer in $SE$, $EN$ and $MO$ platforms. Lowest percentage ($\sim$79\%) of questions have been answered in the $CH$ platform. Further $SE$, $GD$ and $MA$ have a higher percentage of questions having accepted answers than other platforms. The average number of votes to the questions are maximum for the $SE$ platform.\\
\noindent \textbf{Length of the title and body}: We have computed length of the title and body as the number of words present in the each of them respectively. Further we took top 20\% questions ($Q_{top}$) and bottom 20\% questions ($Q_{bottom}$) based on the response time. Top 20\% questions correspond to a lower response time and bottom 20\% questions correspond to a higher response time. In $MA$, $EN$ and $SE$, the average length of the body of $Q_{top}$ questions are 69.06, 71.86, 150.70 respectively; the length of the body of $Q_{bottom}$ questions are 117.48, 102.55, 189.70 respectively. \textit{Thus questions with lesser length get faster answers}. %It is observed that the questions with lesser length of body get fast answers than the questions with larger length of body.\am{you said just the opposite in intro!}~\rh{[fixed issue]} 
The same trend is observed for $MO$, $GD$ and $CH$ where the length of the body for $Q_{top}$ are 121.56, 133.63, 81.197 and for $Q_{bottom}$ are 138.48, 166.15, 106.117 respectively. %So, for all the Q\&A communities, we are observing the similar trend. 
We did not find much difference in the length of the titles for the $Q_{top}$ vs the $Q_{bottom}$ questions.\\
\noindent \textbf{Linguistic characteristics}: In order to understand the linguistic characteristics, we have used three measures -- Flesch reading ease test measure~\footnote{https://simple.wikipedia.org/wiki/Flesch\_Reading\_Ease} (higher the better), Coleman Liau Index~\footnote{https://en.wikipedia.org/wiki/Coleman-Liau\_index} (lower the better) and Automated Readability Index~\footnote{https://en.wikipedia.org/wiki/Automated\_readability\_index}(lower the better). We observe the readability scores for $Q_{top}$ and $Q_{bottom}$ questions.
For Flesch-reading-ease test (see Figure~\ref{fig:readabilityGraphprop}), the $Q_{top}$ questions of $MA$, $EN$ and $SE$ have average scores of 68.53, 71.71, 63.31 respectively while the $Q_{bottom}$ questions have the average scores of 60.59, 68.86 and 58.14 respectively. In case of $MO$, $GD$ and $CH$, the average scores of $Q_{top}$ are 71.42, 62.75 and 63.22 respectively while that of $Q_{bottom}$ questions are 68.83, 55.26 and 58.87 respectively. Thus readability is better (easier) for the $Q_{top}$ questions across all platforms irrespective of their popularity (in terms of the answering speed). This observation holds true for the other two readability metrics as well (see Figure~\ref{fig:readabilityGraphprop}). %Accordingly to Flesch reading ease score,  are easy to understand than $Q_{bottom}$ questions for all the platforms. According to Coleman Liau Index(see figure~\ref{fig:readabilityGraphprop}), the average scores of $Q_{top}$ questions of $MA$, $EN$ and $SE$ are 7.93, 7.76 and 9.40 respectively. Similarly the average scores of $Q_{bottom}$ questions are 9.78, 8.37 and 10.60 respectively. For $MO$, $GD$ and $CH$, the average score of $Q_{top}$ questions are 7.19, 10.13 and 9.30 respectively. Similarly the average score of $Q_{bottom}$ questions are 7.73, 12.55 and 10.21 respectively. According to Coleman Liau Index, $Q_{top}$ questions are easy to understand than the $Q_{bottom}$ questions for all the platforms. For automated readability score (see figure~\ref{fig:readabilityGraphprop}), $MA$, $EN$ and $SE$, the average score of $Q_{top}$ are 12.94, 8.58 and 10.93 respectively. The average score of $Q_{bottom}$ questions are 15.10, 9.48 and 12.45. For $MO$, $GD$ and $CH$, the average score of $Q_{top}$ are 8.61, 11.45 and 10.86 respectively. Similarly, the average score for $Q_{bottom}$ questions are 9.15, 13.72 and 12.11 respectively.}

\noindent \textbf{Tag related analysis}: For each platform, we obtain the most frequently occurring tags (top 50) across the questions posted. %In Figure~\ref{fig:tagcloud}, show the top 50 tags based on their occurrence. 
For $SE$, {\em Java}, {\em Design}, {\em C\#} are the most popular tags. For $GD$, {\em Unity}, {\em C++}, {\em C\#} and {\em OpenGL} are most frequent and for $MO$, {\em United States} and {\em taxes} are most popular tags. In case of $MA$, $EN$ and $CH$, we do not observe any particularly dominant set of tags. Further, for each portal, we calculate the average number of tags present in $Q_{top}$ and $Q_{bottom}$ questions. In $MA$, $EN$ and $SE$, the average number of tags in $Q_{top}$ and $Q_{bottom}$ are 2.08, 2.03, 2.53, 2.66, 2.17 and 3.02 respectively. In $GD$, $CH$ and $MO$, the average number of tags for $Q_{top}$ and $Q_{bottom}$ are 2.58, 2.28, 3.00, 2.94, 2.54 and 3.23 respectively. We observe that there are lesser number of tags in the questions in $Q_{top}$ than in $Q_{bottom}$. This could possibly indicate lesser tags makes a question topically less confusing and streamlines it more effectively to people who can actually answer them.%; hence the observation.\am{The last line makes no sense. I believe that lesser tags makes the question topically less confusing and streamlines it more effectively to people who can actually answer them; hence the observation.}~\rh{[updated]}
% \begin{figure*}[!ht]
% %\centering
% \includegraphics[height=3.5cm, width=5.4cm]{Figures/math_wordcloud.png}\hfill
% \includegraphics[height=3.5cm, width=5.4cm]{Figures/english_wordcloud.png}\hfill
% \includegraphics[height=3.5cm, width=5.4cm]{Figures/software_eng_wordcloud.png}\hfill\\
% %\caption{(Math, English, Software Engineering) tags wordcloud}
% %\centering
% \includegraphics[height=3.5cm, width=5.4cm]{Figures/chemistry_wordcloud.png}\hfill
% \includegraphics[height=3.5cm, width=5.4cm]{Figures/gamedev_wordcloud.png}\hfill
% \includegraphics[height=3.5cm, width=5.4cm]{Figures/money_wordcloud.png}
% \caption{(Chemistry, GameDev, Money) tags wordcloud}
% \label{fig:tagcloud}
% \end{figure*}

\noindent \textbf{Votes received by the question}: In order to understand the temporal pattern of votes received by the questions, we count the fraction of questions receiving 1 upvote, 2 upvotes and $\ge 3$ upvotes on three consecutive days, i.e., the same day when the question is posted followed by the two subsequent days. In the same day, $\sim$30\% questions of $SE$ platform receive more than 3 votes. But for other platforms most of the questions received one vote. In second day, around $\sim$20\% of questions receive one more vote in $SE$, $MO$ and $CH$ platforms. In case of $SE$ and $MO$, there are 10\% of questions receiving at least three votes in the second day. In the third day, there are very less number of questions receiving votes across all the platform. This indicates that on all platforms the attention span of users for a particular question dies down within 48 hours from the post of the question. In case of $MA$, $EN$ and $SE$, the average votes for the questions in $Q_{top}$ are 1.99, 5.31 and 15.82 respectively and that for the questions in $Q_{bottom}$ are 2.75, 2.60 and 4.44 respectively. In case of $MO$, $GD$ and $CH$, the average number of votes in $Q_{top}$ are 6.46, 4.63 and 4.21 respectively while that in $Q_{bottom}$ are 2.96, 2.08 and 5.17 respectively. Hence questions in lower response time typically receive more upvotes irrespective of the popularity of the platform. %Also, we have computed the average number of votes received by the questions within 24 hours of posting the question. We observe the similar trend for all the platforms. \am{How is the behavior in $Q_{top}$ and $Q_{bottom}$?}~\rh{added}
\subsection{Characterizing the users}
In this section we analyse the user behaviour on each of these platforms. Number of users in $MA$, $SE$ and $EN$ are approximately 888K, 346K and 350K which is much larger than the number of users in other platforms. For $MO$, $GD$ and $CH$, the number of users are 84K, 122K and 88K respectively. In Table~\ref{table:userAnalysis}, we show the basic statistics about these unique users such as %total number of unique users \am{not there in table}, 
\% of users only posting question, \% of users only answering questions, \% of users posting both the question and answers, and users whose at least one answer got accepted by an asker. 
%\vspace*{-0.65cm}
\begin{table}[htbp]
  %\centering
  \tiny
\resizebox{\textwidth}{!}{%  
    \begin{tabular}{|c|c|c|c|c|}
    \hline
    Datasets & \%Users posting only questions & \%Users posting only answers & \%Users posting both Q\&A & \%Users whose at least one answer got accepted    \\ 
     \hline
     \em{SE} & 8.9  & 7.5 & 2 & 1.9  \\ \hline
     \em{GD} & 19 & 12.7 & 5.3 & 4.9 \\ \hline
     \em{MA} & 33 & 10 & 5 & 4 \\ \hline
     \em{EN} & 17 & 12.3 & 2.6 & 2.3 \\ \hline
    \em{CH} & 20 & 7 & 2.6 & 2.1 \\ \hline
     \em{MO} & 22.2 & 9.8 & 2.7 & 2.2 \\ 
     \hline
    \end{tabular}%
    }
    \caption{Basic statistics of the user interaction.}
\label{table:userAnalysis}
%\vspace*{-0.85cm}
\end{table}

We have seen that there are around 60\%-70\% users who remain inactive in all the platforms. Inactive users neither post anything, nor do they interact with anyone on the platform. $MA$, $SE$ and $EN$ platforms have approximately 37.38\%, 14.14\% and 26.73\% of users who have posted at least one question or one answer (aka active users). In case of $MO$, $GD$ and $CH$, the percentage of active users are 29.38\%, 26.47\% and 25.29\% respectively. According to Table~\ref{table:userAnalysis}, $MA$ has maximum fraction of users (33\% users) who have posted at least one question. In terms of writing answers, around 12\% users are active in $EN$ and $GD$ platforms which is larger than the other platforms. $GD$ has maximum fraction of users ($\sim$5.3\%) who have posted both question and answer. Once again, $GD$ has the largest fraction of users (~4.9\%) whose at least one answer got selected. Further we conduct in depth analysis on the users who have asked or answered any questions in $Q_{top}$ and $Q_{bottom}$ respectively. In Table~\ref{table:Asker and  Answerer behaviour}, we show number of askers, answerers and their overlap for $Q_{top}$ and $Q_{bottom}$ users. We analyze various factors of the asker and answerer of the questions in $Q_{top}$ and $Q_{bottom}$ below.

%\am{You need to connect this again $Q_{top}$ and $Q_{bottom}$. Take these two segments of questions and analyze their asker-answerer behaviour. That would make sense.}~\rh{[added]}
\noindent \textbf{Reputation score of users}: We separately observe the reputation scores of asker and answerer of questions of $Q_{top}$ and $Q_{bottom}$ category. Asker is the user who asked the question in the portal and answerer is the user who answered the question. In $MA$, $EN$ and $SE$, the average reputation scores of askers in $Q_{top}$ are 559.07, 546.27 and 1001.96 respectively. The average reputation scores of askers in $Q_{bottom}$ 705.52, 522.78 and 644.14 respectively. In $MO$, $GD$ and $CH$, the average reputation score of askers in $Q_{top}$ are 585.71, 401.53 and 424.68 respectively. The average reputation score of askers in $Q_{bottom}$ are 540.71, 307.14 and 561.10 respectively. In case of answerers, the average reputations in $Q_{top}$ are 4063.39, 2147.42 and 3209.59 for $MA$, $EN$ and $SE$ respectively and those in $Q_{bottom}$ are 2001.24, 1622.79 and 1846.81 respectively. For $MO$, $GD$ and $CH$, the average reputation of answerer in $Q_{top}$ are 4017.05, 1117.52 and 1948.33 respectively while that in $Q_{bottom}$ are 1965.42, 552.84 and 1397.38 respectively. We observe three things here -- (i) across all platforms the reputation scores of askers as well as answerers are typically more for $Q_{top}$ questions (ii) in the popular platforms the reputation scores for both askers and answerers are higher and (iii) answerer reputations are an order of magnitude higher than the askers.
%We consider top 20\% and bottom 20\% users based on their reputation scores to observe what is the average number of questions they have answered, average number of answers they have posted. In $MA$, the average number of questions answered by top 20\% users is 11.14 whereas the average of number of questions answered by bottom 20\% users is 1.25. For $CH$,  $EN$ and $GD$, the average number of questions answered by top 20\% users are 3.61, 3.02, 3.21 respectively. For $MO$ and $SE$, the value for top 20\% users are 2.55 and 2.19 respectively. For bottom 20\% users in all the platforms, the average number of questions answered lies in between 1.03--1.56.\\

\noindent\textbf{Asker-Answerer graph (AAG)}: %\am{Again here, build two AAGs corresponding to the users in $Q_{top}$ and $Q_{bottom}$. Compare their network properties.} 
To understand the user interaction patterns on a platform, we build an undirected graph where the nodes are the users and the edges denote if a user (answerer) answers a question posted by another user (asker). 
%In table~\ref{table:2}, we have presented basic statistics such as number of nodes, number of edges in largest connected components, number of connected components and clustering coefficient of asker-answerer graph for each platform. In Table ~\ref{table:2}, it is observed that for $MA$ and $MO$, the clustering coefficient of the network is relatively higher than the other platforms. So, Users in $MA$ and $MO$ platforms are more densely connected. The clustering coefficient of asker-answerer graph of $SE$, $GD$ are quite low (~0.03) which indicate that the users are not well connected among themselves.\\
For every platform, we build two such AAGs for users who either asked or answered the questions of $Q_{top}$ and $Q_{bottom}$ respectively (See Table~\ref{table:Comparison of network properties}). Further, we calculate average degree, average clustering coefficient for the whole network. In Figure~\ref{fig:readabilityGraphprop} (e), we observe that the average degree of $Q_{top}$ AAG for all platforms except $CH$ are relatively higher than the average degree of $Q_{bottom}$ AAG. For all the platforms, the average clustering coefficient of $Q_{top}$ AAG are higher than the average clustering coefficient of $Q_{bottom}$ AAG (see Figure~\ref{fig:readabilityGraphprop} (d)). This indicates that there is a higher density of interactions among the asker-answer(s) in the $Q_{top}$ AAG (and exceptionally higher platforms $SE$, $GD$ and $MO$). %But in case of $SE$, $GD$ and $MO$, the average clustering for $Q_{top}$ AAG much higher than $Q_{bottom}$ AAG which infers that the nodes of $Q_{top}$ AAG are more closely connected among each other than the nodes in $Q_{bottom}$ AAG. We show the average clustering coefficient for $Q_{top}$ and $Q_{bottom}$ AAG for all the platforms in 
%we observe that there are common users present across these six platforms. So, we investigate the various behaviour of these users in the next section.\\
%\vspace*{-0.75cm}
\begin{table}[ht]
\tiny
\resizebox{\textwidth}{!}{
    % \centering
    \begin{tabular}{|c|c|c|c|} 
     \hline
     Datasets & \# nodes & \# edges   & fraction of nodes present in largest connected component.\\ 
     \hline
     \em{SE} ($Q_{top}, Q_{bottom}$) & 10319, 12756 & 11179, 11234 & 0.73, 0.53 \\  \hline
     % \em{SE} ($Q_{bottom}$) & 12756 & 11234 & 0.53 \\ 
     \em{GD} ($Q_{top}, Q_{bottom}$)  & 8011, 9453 & 8730, 8547 & 0.72, 0.42 \\ \hline
     % \em{GD} ($Q_{bottom}$) & 9453 & 8547 & 0.42 \\ 
     \em{MA} ($Q_{top}, Q_{bottom}$)  & 106948, 110284 & 218119, 210865 & 0.92, 0.865 \\ \hline
     % \em{MA} ($Q_{bottom}$) & 110284  & 210865 & 0.865 \\ 
     \em{EN} ($Q_{top}, Q_{bottom}$)  & 19609, 20672 & 21499, 21800 & 0.702, 0.71 \\ \hline
     % \em{EN} ($Q_{bottom}$) & 20672  & 21800 & 0.71 \\ 
     \em{CH} ($Q_{top}, Q_{bottom}$)  & 5350, 5336 & 5953, 6069 & 0.79, 0.73 \\ \hline
     % \em{CH} ($Q_{bottom}$) & 5336  & 6069 & 0.73 \\ 
     \em{MO} ($Q_{top},Q_{bottom}$)  & 5093, 6041 & 5966, 6096 & 0.87, 0.69 \\
     % \em{MO} ($Q_{bottom}$) & 6041  & 6096 & 0.69 \\  [1ex] 
     \hline
    \end{tabular}
    }
\caption{Comparison of network properties of $Q_{top}$ \& $Q_{bottom}$ AAGs.}
\label{table:Comparison of network properties}
%\vspace*{-0.85cm}
\end{table} 
%\vspace*{-0.55cm}
\begin{table}[ht]
\resizebox{\textwidth}{!}{
    \begin{tabular}{|c|c|c|c|c|c|c|} 
     \hline
     Datasets & Number of askers in $Q_{top}$ &  Number of answerer in $Q_{top}$   & Number of asker in $Q_{bottom}$ & Number of answerer in $Q_{bottom}$ & overlap of askers & overlap of answerers\\ 
     \hline
     \em{SE} & 7697 & 2861 & 8933 & 4701 & 797 & 878\\  \hline
     \em{GD} & 5959 & 2249 & 6417 & 4941 & 843 & 1905\\ \hline
     \em{MA}   & 89557 & 14072 & 88772 & 39909 & 6891 & 18397 \\ \hline
     \em{EN}  & 14533 & 4542 & 15564  & 6349 & 1148 & 1241\\ \hline
     \em{CH}  & 4408 & 1180 & 4021 & 1860 & 427 & 545\\ \hline
     \em{MO}  & 4497 & 762 & 4724 & 1789 & 256 & 472\\
     \hline
    \end{tabular}
}
\caption{Asker and Answerer behaviour $Q_{top}$ \& $Q_{bottom}$}
\label{table:Asker and  Answerer behaviour}
%\vspace*{-0.85cm}
\end{table} 
\noindent\textbf{Cross platform users}: %\am{Again we should analyze the common users in $Q_{top}$ across platforms vs common users in $Q_{bottom}$ across platforms.}
We found around 2739 users who are present in all the six platforms. %So, we attempt to find out how many of them are actually active in each platform. In order to find this, we first check fraction of common users who has posted atleast one question then we observe the fraction of common users who have posted atleast one answer.
Fraction of common users posted at least one question is around 29\%-30\% for $MA$ and $EN$ and for $SE$ it is 18\%. However for $CH$, $GD$ and $MO$ platforms, this fraction comparatively lower ($\le 15\%$). The observations are similar for the number of common users posting an answer -- the popular platforms have far larger fractions compared to less popular ones. %While looking at the fraction of common users posted at least one answer, we observe that the fraction is quite high for $MA$, $SE$ and $EN$. But in case of $CH$, $GD$ and $MO$, the fraction remain low (in the range of 6\%-13.3\%).\\
%Next, we observe what fraction of common users are most active in which platform. For every user, we look at the number of questions and answers posted to each platform. User is considered as most active in a particular portal if he has posted maximum number of questions and answers in the portal than the other portals. We found that 19\% of the common users are most active in $EN$ portal and ~15\% of common users are most active in $MA$ platform. In $SE$, 10\% common users are most active. In case of $CH$, $GD$ and $MO$, percentage of most active common users are 2.37\%, 4.6\% and 5\% respectively. We observe there are around ~42\% common users who remain inactive in all the platforms. Further we observe what fraction of answers are actually posted by the common users. $MO$ and $SE$ are the platforms where the common users have posted around ~7.5\% of the answers. But in case of $GD$, $MA$ and $CH$, only 2.17\%-5.11\% answers are posted by common users.\\
Next we compute the number of common users who participated in asking/answering a question in $Q_{top}$ and $Q_{bottom}$. %We compute the fraction of the total number of common users who answer any question in a category ($Q_{top}$, $Q_{bottom}$). 
The percentage of common users asking one or more questions in case of $MA$, $EN$ and $SE$ are 19.82\%, 17.30\% and 9.71\% respectively in $Q_{top}$ and 20.0\%, 16.24\% and 9.49\% respectively in $Q_{bottom}$. The percentage of common users in these three platforms who answered one or more questions in $Q_{top}$ are 9.3\%, 8.72\% and 6.17\% respectively. In $Q_{bottom}$ these percentages are 12.92\%, 7.7\% and 6.4\% respectively. For $MO$, $GD$ and $CH$, the percentage of common users acting as asker in $Q_top$ are 7.84\%, 6.38\% and 5.69\% respectively and in $Q_{bottom}$ are 8.06\%, 6.09\% and 5.69\% respectively.  For these three platforms, the percentage of common users acting as answerer in $Q_top$ are 2.7\%, 4.3\% and 1.8\% respectively and in $Q_{bottom}$ are 4.2\%, 4.9\% and 2.1\% respectively.\\

\noindent \textbf{Engagement of common users}: In order to investigate the involvement of common users (present in $Q_{top}$ and $Q_{bottom}$ as answerers) in the overall platform, we  compute the average number of questions/answers posted by them at an aggregate level. For $MA$, $EN$ and $SE$, the average number of questions posted by common users in $Q_{top}$ are 22.68, 8.5 and 3.04 and that in $Q_{bottom}$ are 21.60, 8.74 and 3.57 respectively. For these three platforms, the average number of answers posted by common users in $Q_{top}$ are 228.83, 47.32 and 61.01 and that in $Q_{bottom}$ are 166.7, 52.20 and 56.74 respectively. For $MO$, $GD$ and $CH$, the average number of questions posted by common users in $Q_{top}$ are 11.56, 5.1 and 5.61 respectively. and in $Q_{bottom}$ are 9.02, 5.45 and 7.55 respectively. For these three platforms, the average number of answers posted by common users in $Q_{top}$ are 49.81, 24.74 and 15.55 respectively. The average number of answers posted by common users in $Q_{bottom}$ are 35.11, 21.43 and 14.22 respectively.
Thus (i) irrespective of the popularity of the platforms, common users engage more in terms of asking and answering in $Q_{top}$ than in $Q_{bottom}$ and (ii) quite interestingly, the average number of answers posted by the common users in $Q_{top}$ and $Q_{bottom}$ are far more than the questions posted by them which indicates that these users are more engaged in posting answers than questions. 
\section{Question Category Prediction}
%\vspace*{-0.2cm}
\subsection{Classification model}
%\vspace*{-0.1cm}
We consider previously discussed properties of the questions as features and predict whether a question will belong to $Q_{top}$ or $Q_{bottom}$. The features are 
 -- length of question body, Flesch reading ease, Coleman Liau index, automated readability index, number of tags, reputation of the asker and the clustering coefficient of asker in the asker-answerer graph.  For every dataset, we have randomly divided the whole dataset into three parts -- training (70\%), validation (10\%), and test (20\%). Given a dataset, the number of instances present in training, validation, and test are given in Table~\ref{tab:TrVlTs}. For this experiment, we have used four standard machine learning classifiers -- (a) logistic regression (LR), (b) support vector classifier (SVC), (c) random forest (RF) (d) XGBoost. For evaluation, we have used overall precision, recall and macro F1 score.\\
 \begin{table}[!h]
\centering
\tiny
\scalebox{0.99}{
\begin{tabular}{|c|c|c|c|}
\hline
\textbf{Datasets} & \textbf{Train Size} & \textbf{Valid Size} & \textbf{Test Size} \\ \hline
${SE}$                & 15848               & 2268                & 4525               \\ \hline
${GD}$                & 12660               & 1812                & 3615               \\ \hline
${MA}$                & 325153              & 46543               & 92809              \\ \hline
${EN}$                & 31119               & 4454                & 8884               \\ \hline
${CH}$                & 8844                & 1266                & 2525               \\ \hline
${MO}$                & 8875                & 1270                & 2534               \\ \hline
\end{tabular}
}
\caption{Number of instances present in training, validation and testing for all the datasets}
\label{tab:TrVlTs}
%\vspace*{-0.85cm}
\end{table}
\noindent \textbf{Parameter settings}: The parameter settings for each model are given below. For SVC, we consider the value of parameter $C$ in the range of 1-3. The values tried for kernel are poly, rbf and sigmoid. For random forest classifier, the range of the number of estimators is between 160 to 200. The criterion values are gini and entropy. For XGBoost, the range of values for the number of estimators is from 160 to 200. The range of learning rate is 0.1 to 0.3.

\noindent \textit{Logistic regression (LR)}: For all datasets, we found the default parameters are the best.\\
\noindent \textit{Support vector classifier (SVC)}: For $CH$, the best value of $C$ is 3, and kernel is set to {\em poly}. For other datasets, the best parameter for {\em kernel} is rbf. For $EN$ and $GD$, the value of the parameter {\em C} is set to 2 and 1 respectively. For the other datasets, the value of {\em C} is 3.\\
\noindent \textit{Random forest (RF)}: For $CH$, $EN$, $MO$, $MA$, the best value for criterion is {\em entropy}. For $GD$ and $SE$, the criterion is set to {\em gini}. For $MO$, the number of estimators is set to 180. For other datasets, the number of estimators is set to 200.\\
\noindent \textit{XGBoost}: For all the datasets, the best value for the learning rate is 0.1. For $CH$, $GD$, $MO$, the value of the number of estimators is set to 160. For $EN$ and $SE$, the number of estimators is 180. For $MA$, the number of estimators is 200.

\subsection{Results}
%\vspace*{-0.85cm}
\begin{table*}[ht]%[!ht]
\centering
\resizebox{\textwidth}{!}{
\begin{tabular}{|c|ccc|ccc|ccc|ccc|}
\hline
\multirow{2}{*}{\textbf{Datasets}} & \multicolumn{3}{c|}{\textbf{LR}}                                                  & \multicolumn{3}{c|}{\textbf{SVC}}                                                                  & \multicolumn{3}{c|}{\textbf{RF}}                                             & \multicolumn{3}{c|}{\textbf{XGBoost}}                                                              \\ \cline{2-13} 
                                   & \multicolumn{1}{c|}{\textbf{Precision}} & \multicolumn{1}{c|}{\textbf{Recall}} & \textbf{Macro F1} & \multicolumn{1}{c|}{\textbf{Precision}} & \multicolumn{1}{c|}{\textbf{Recall}} & \textbf{Macro F1} & \multicolumn{1}{c|}{\textbf{Precision}} & \multicolumn{1}{c|}{\textbf{Recall}} & \textbf{Macro F1} & \multicolumn{1}{c|}{\textbf{Precision}} & \multicolumn{1}{c|}{\textbf{Recall}} & \textbf{Macro F1} \\ \hline
\textbf{${SE}$}                        & \multicolumn{1}{c|}{0.67}               & \multicolumn{1}{c|}{0.67}            & 0.67              & \multicolumn{1}{c|}{0.69}               & \multicolumn{1}{c|}{0.69}            & 0.69              & \multicolumn{1}{c|}{0.71}               & \multicolumn{1}{c|}{0.71}            & \underline{0.71}              & \multicolumn{1}{c|}{0.72}               & \multicolumn{1}{c|}{0.72}            & \textbf{0.72}              \\ \hline
\textbf{${GD}$}                        & \multicolumn{1}{c|}{0.62}               & \multicolumn{1}{c|}{0.61}            & 0.61              & \multicolumn{1}{c|}{0.63}               & \multicolumn{1}{c|}{0.62}            & \underline{0.62}              & \multicolumn{1}{c|}{0.62}               & \multicolumn{1}{c|}{0.62}            & \underline{0.62}              & \multicolumn{1}{c|}{0.63}               & \multicolumn{1}{c|}{0.63}            & \textbf{0.63}              \\ \hline
\textbf{${MA}$}                        & \multicolumn{1}{c|}{0.69}               & \multicolumn{1}{c|}{0.69}            & 0.69              & \multicolumn{1}{c|}{0.70}               & \multicolumn{1}{c|}{0.70}            & \underline{0.70}              & \multicolumn{1}{c|}{0.69}               & \multicolumn{1}{c|}{0.69}            & 0.69              & \multicolumn{1}{c|}{0.71}               & \multicolumn{1}{c|}{0.71}            & \textbf{0.71}              \\ \hline
\textbf{${EN}$}                        & \multicolumn{1}{c|}{0.61}               & \multicolumn{1}{c|}{0.61}            & 0.60              & \multicolumn{1}{c|}{0.61}               & \multicolumn{1}{c|}{0.61}            & 0.61              & \multicolumn{1}{c|}{0.63}               & \multicolumn{1}{c|}{0.63}            & \underline{0.63}              & \multicolumn{1}{c|}{0.64}               & \multicolumn{1}{c|}{0.64}            & \textbf{0.64}              \\ \hline
\textbf{${CH}$}                        & \multicolumn{1}{c|}{0.59}               & \multicolumn{1}{c|}{0.58}            & \underline{0.58}              & \multicolumn{1}{c|}{0.59}               & \multicolumn{1}{c|}{0.58}            & 0.57              & \multicolumn{1}{c|}{0.59}               & \multicolumn{1}{c|}{0.58}            & \textbf{0.59}              & \multicolumn{1}{c|}{0.58}               & \multicolumn{1}{c|}{0.58}            & \underline{0.58}              \\ \hline
\textbf{${MO}$}                        & \multicolumn{1}{c|}{0.56}               & \multicolumn{1}{c|}{0.56}            & \underline{0.56}              & \multicolumn{1}{c|}{0.56}               & \multicolumn{1}{c|}{0.56}            & \underline{0.56}              & \multicolumn{1}{c|}{0.56}               & \multicolumn{1}{c|}{0.56}            & \underline{0.56}              & \multicolumn{1}{c|}{0.57}               & \multicolumn{1}{c|}{0.57}          & \textbf{0.57}              \\ \hline
\end{tabular}
}
\caption{Results of question classification models for all the datasets. The best results are highlighted in \textbf{bold}. The $2^{nd}$ best results are highlighted in \underline{underline.}}
\label{tab:result}
%\vspace*{-0.85cm}
\end{table*}
In Table~\ref{tab:result}, we present the overall precision, recall and macro F1 score for all the datasets. For $SE$, RF and XGBoost are performing better than other models. For $GD$, almost all the models' performances are similar (0.61-0.63 F1 score). For $MA$, XGBoost model is performing better (0.71 macro F1 score) than other models. For $EN$, RF and XGBoost attain better F1 score than other models. For $CH$, RF performs better (0.59 F1 score) than other models. For $MO$, all the models attain similar performance (0.56-0.57 F1 score). %\am{are these numbers good?}
\section{Conclusion}
In this paper, we study question-answering trends across six diverse CQA platforms. We find metadata, question structure, and user interaction patterns affect response time. Shorter, clearer questions with fewer tags are answered faster. High-reputation users engage more with quickly-answered questions. These question and asker features predict fast responses. We use machine learning to classify new questions based on metadata, planning to incorporate these factors into deep learning models to predict response times for new questions using text, metadata, and asker features.
% In this paper, we examine question-answering trends spanning six disparate platforms. We discover that factors such as metadata, the way questions are structured, and user interaction patterns exert influence on the response time. Questions that are shorter and clearer with fewer tags tend to receive faster answers. Users with a high reputation tend to engage more with questions that are quickly answered. These question characteristics and the attributes of the person asking the question can be used to predict rapid responses. Using machine learning, we classify new questions based on metadata, with plans to integrate these factors into advanced deep learning models. This integration will enable the prediction of response times for new questions by using text, metadata, and characteristics of the person asking the question.

\bibliographystyle{plainnat}
\bibliography{references}

% \section*{References}

% References follow the acknowledgments. Use unnumbered first-level heading for
% the references. Any choice of citation style is acceptable as long as you are
% consistent. It is permissible to reduce the font size to \verb+small+ (9 point)
% when listing the references.
% {\bf Note that the Reference section does not count towards the eight pages of content that are allowed.}
% \medskip

% \small

% [1] Alexander, J.A.\ \& Mozer, M.C.\ (1995) Template-based algorithms for
% connectionist rule extraction. In G.\ Tesauro, D.S.\ Touretzky and T.K.\ Leen
% (eds.), {\it Advances in Neural Information Processing Systems 7},
% pp.\ 609--616. Cambridge, MA: MIT Press.

% [2] Bower, J.M.\ \& Beeman, D.\ (1995) {\it The Book of GENESIS: Exploring
%   Realistic Neural Models with the GEneral NEural SImulation System.}  New York:
% TELOS/Springer--Verlag.

% [3] Hasselmo, M.E., Schnell, E.\ \& Barkai, E.\ (1995) Dynamics of learning and
% recall at excitatory recurrent synapses and cholinergic modulation in rat
% hippocampal region CA3. {\it Journal of Neuroscience} {\bf 15}(7):5249-5262.

\end{document}